\documentclass[aps,prl,twocolumn]{revtex4}

\usepackage{amsmath}
\usepackage{graphicx}
\usepackage{bm}
\DeclareGraphicsExtensions{.ps}

\preprint{JLAB-THY-12-1577}

\begin{document}

\title{Comment on ``Taming the Pion Cloud of the Nucleon''}

\author{Chueng-Ryong Ji$^{\, a}$, W. Melnitchouk$^b$, A. W. Thomas$^c$}
\affiliation{
   $^a$ Department of Physics, % Box 8202,
	North Carolina State University,
	Raleigh, North Carolina 27692, USA	\\
   $^b$ Jefferson Lab, 12000 Jefferson Avenue,
        Newport News, Virginia 23606, USA	\\
   $^c$ \mbox{CSSM and CoEPP, School of Chemistry and Physics,
	University of Adelaide, Adelaide SA 5005, Australia}}

\maketitle

%%%%%%%%%%%%%%%%%%%%%%%%%%%%%%%%%%%%%%%%%%%%%%%%%%%%%%%%%%%%%%%%%%%%%%%%%
In a recent Letter, Alberg and Miller (AM) \cite{Alberg12} presented a
calculation of the pion contributions to the self-energy of the nucleon,
in an attempt to better constrain the role of the pion cloud in the
$\bar d - \bar u$ asymmtry in the proton sea.
The self-energy $\Sigma$ was computed from a pseudoscalar (PS)
pion--nucleon interaction
($\bar\psi_N \gamma_5 \vec\tau \psi_N \cdot \vec\phi_\pi$),
with the claim that the result would be equivalent to that with
the more usual pseudovector (PV) form
($\bar\psi_N \gamma_\mu \gamma_5 \vec\tau \psi_N \cdot
  \partial^\mu \vec\phi_\pi$).
The PV theory is consistent with the chiral symmetry properties of the
strong interactions, as embodied for example in chiral perturbation
theory, whereas the PS coupling requires in addition a scalar field to
restore chiral invariance \cite{Weinberg67}.

In this Comment we demonstrate that the PV and PS pion--nucleon
couplings do in fact lead to different results for the self-energy.
In particular, for the model-independent, leading nonanalytic (LNA)
behavior of the self-energy, the PV theory yields the well-known
$m_\pi^3$ dependence in the chiral limit, while the PS interaction
involves an additional, lower-order term $\sim m_\pi^2 \log m_\pi^2$
\cite{Ji09}.
To identify the origin of the difference, we can express the total
self-energy for the PS coupling $\Sigma^{\rm PS}$ in terms of the PV
self-energy $\Sigma^{\rm PV}$ and a contribution from the end-point
region corresponding to $k^+=0$,
$\Sigma^{\rm PS} = \Sigma^{\rm PV} + \Sigma^{\rm PS}_{\rm end-pt}$.
Using the Goldberger-Treiman relation to relate the $\pi NN$ coupling
to the axial charge of the nucleon $g_A$ and the pion decay constant
$f_\pi$, the end-point contribution can be written as
\begin{equation}
\Sigma^{\rm PS}_{\rm end-pt}
= \frac{3 i g_A^2 M}{2 f_\pi^2}
  \int\!\!\frac{d^4 k}{(2\pi)^4}
  \frac{F^2(k^2,(p-k)^2)}{k^2-m^2_\pi+i\epsilon},
\label{eq.1}
\end{equation}
where, in analogy with AM, we introduce a $\pi NN$ form factor $F$
which suppresses contributions from short distances.  Although our
results do not depend on the details of the short-distance $\pi N$
interaction, for generality we keep the dependence of the form factor
on the invariant mass of both the intermediate state pion and nucleon.

When performing the $k^-$ integration, it is crucial to realize
that the pion pole depends not only on the sign of $k^+$, but also
effectively runs to infinity as $k^+ \to 0$.
Keeping this runaway pole in the $k^-$ contour integration is the
key to the difference between $\Sigma^{\rm PS}$ and $\Sigma^{\rm PV}$.
In Ref.~\cite{Alberg12}, AM consider only $k^+>0$ and $k^+<0$,
and omit the contribution from the end-point $k^+=0$.
Including the pole at infinity, one finds \cite{Ji09}
\begin{equation}
\label{eq.2}
\Sigma^{\rm PS}_{\rm end-pt}
% = {3 g_A^2 M^2 \pi}{8M f_\pi^2 (2\pi)^3} \int^\infty_0 dt
% \frac{4\sqrt{t}\, F^2(m_\pi^2,-t)}{\sqrt{t+m_\pi^2}},
= \frac{3 g_A^2 M}{16 \pi^2 f_\pi^2} \int^\infty_0 dt
  \frac{\sqrt{t}\, F^2(m_\pi^2,-t)}{\sqrt{t+m_\pi^2}},
\end{equation} 
where the form factor is evaluated at the pion pole.

To evaluate the contribution in Eq.~(\ref{eq.2}) explicitly,
we can use a dipole parametrization for the dependence of the
form factor on the nucleon virtuality,
$F(m_\pi^2,-t)=((\Lambda^2 - M^2)/(\Lambda^2 + t))^2$,
with $\Lambda$ a mass parameter.
Following AM, we define $a=m_\pi^2/M^2$, $b=\Lambda^2/M^2$
and find
\begin{eqnarray}
\Sigma^{\rm PS}_{\rm end-pt}
% &=& \frac{3 \pi g^2_{\pi N}}{8M(2\pi)^3} 
&=& \frac{3 g_A^2 M}{64 \pi^2 f_\pi^2}
\Big\{ \sqrt{b(a-b)}(a-4b)(3a+2b)		\nonumber\\
& & \hspace{-2cm}
     +\, 3a(a^2-4ab+8b^2)\tan^{-1}\sqrt{\frac{a}{b}-1}
\Big\}
\frac{(b-1)^4}{6(a-b)^{\frac{7}{2}}b^{\frac{5}{2}}}.
\label{eq.4}
\end{eqnarray}
Expanding the term proportional to ${\rm tan}^{-1}\sqrt{a/b-1}$
about $a=0$, the LNA term is found to be $\sim a\log{a}$, which is
of lower order than the LNA term for the PV coupling $\sim m_\pi^3$.
The lowest nonanalytic terms for the total PS self-energy are then
given by \cite{Ji09}
\begin{eqnarray}
\label{eq.5}
\Sigma^{\rm PS}_{\rm nonanal.}
&=& {3 g_A^2 \over 32 \pi f_\pi^2}
\Big( {M \over \pi} m_\pi^2 \log m_\pi^2
      - m_\pi^3					\nonumber\\
& & \hspace*{1cm}
  - {m_\pi^4 \over 2\pi M^2} \log{m_\pi^2 \over M^2}
  + {\cal O}(m_\pi^5)
\Big).
\end{eqnarray}
%
% where we have now correct the typo of missing the $\pi$ factor in the
% denominator of the coefficient in front of $\log{m_\pi^2 \over M^2}$.
%
Note that this result is independent of the short-distance part of
the $\pi NN$ interaction (or the form factor $F$), and can be verified
using either light-front, time-ordered or covariant perturbation theory
\cite{Ji09}.

By using the PS theory and omitting the end-point singularities at
$k^+=0$, AM happen to obtain the same result as given by the PV theory.
However, this {\it ansatz} will not give the correct PV result for
other quantities, such as the pion momentum distribution, $f_\pi^N$
\cite{Alberg12,Sullivan72}.
For example, the moment of $f_\pi^N$ (which corresponds to the
pion loop contribution to the vertex renormaliation $Z_1^\pi$)
in the PS theory gives for the LNA term a value 4/3 larger than
for the PV theory \cite{Thomas00,Detmold01}, with the difference
given by an end-point contribution in the PV case.
The pseudoscalar coupling therefore cannot in general be used if one
wishes to ensure consistency with the chiral properties of QCD, which
are respected by the pseudovector $\pi N$ coupling.

\vspace*{0.5cm}

This work was supported by the DOE contract No. DE-AC05-06OR23177,
under which Jefferson Science Associates, LLC operates Jefferson Lab,
DOE contract No. DE-FG02-03ER41260, and the Australian Research Council
through an Australian Laureate Fellowship.
\newpage

% \vspace*{-0.5cm}

%%%%%%%%%%%%%%%%%%%%%%%%%%%%%%%%%%%%%%%%%%%%%%%%%%%%%%%%%%%%%%%%%%%%%%%%%

\end{document}